\documentclass[sigconf]{acmart}

\usepackage{algorithm}
\usepackage{algpseudocode}
\usepackage{balance}
\usepackage{tcolorbox}
\usepackage{bm}
\tcbuselibrary{skins, breakable}
\AtBeginDocument{%
  }

\copyrightyear{2025}
\acmYear{2025}
\setcopyright{cc}
\setcctype{by}
\acmConference[CIKM '25]{Proceedings of the 34th ACM International
Conference on Information and Knowledge Management}{November 10--14,
2025}{Seoul, Republic of Korea}
\acmBooktitle{Proceedings of the 34th ACM International Conference on
Information and Knowledge Management (CIKM '25), November 10--14, 2025,
Seoul, Republic of Korea}\acmDOI{10.1145/3746252.3761214}
\acmISBN{979-8-4007-2040-6/2025/11}

\begin{document}

\title[M-$LLM^3$REC for Recommendation System with LLMs]{M-$LLM^3$REC: A Motivation-Aware User-Item Interaction Framework for Enhancing Recommendation Accuracy with LLMs}

\author{Lining Chen}
\authornote{These authors contributed equally to this work.} 
\orcid{0009-0009-7985-401X}
\affiliation{%
    \department{School of Electrical and Computer Engineering}
    \institution{The University of Sydney}
    \streetaddress{Maze Cres}
    \city{Sydney}
    \state{NSW}
    \postcode{2008}
    \country{Australia}
}
\email{lche8972@uni.sydney.edu.au}                  

\author{Qingwen Zeng}
\authornotemark[1]       
\orcid{0009-0002-7926-3606}
\affiliation{%
    \department{School of Electrical and Computer Engineering}
    \institution{The University of Sydney}
   \streetaddress{Maze Cres}
    \city{Sydney}
    \state{NSW}
    \postcode{2008}
    \country{Australia}
}
\email{qzen5227@uni.sydney.edu.au}

\author{Huaming Chen}
\orcid{0000-0001-5678-472X}
\authornote{Corresponding author.} 
\affiliation{%
    \department{School of Electrical and Computer Engineering}
      \institution{The University of Sydney}
  \streetaddress{Maze Cres}
    \city{Sydney}
    \state{NSW}
    \postcode{2008}
    \country{Australia}
}
\email{huaming.chen@sydney.edu.au}

\renewcommand{\shortauthors}{Lining Chen, Qingwen Zeng, and Huaming Chen}

\begin{abstract}

Recommendation systems have been essential for both user experience and platform efficiency by alleviating information overload and supporting decision-making. Traditional methods, i.e., content-based filtering, collaborative filtering, and deep learning, have achieved impressive results in recommendation systems. However, the cold-start and sparse-data scenarios are still challenging to deal with. Existing solutions either generate pseudo-interaction sequence, which often introduces redundant or noisy signals, or rely heavily on semantic similarity, overlooking dynamic shifts in user motivation. To address these limitations, this paper proposes a novel recommendation framework, termed M-$LLM^3$REC, which leverages large language models for deep motivational signal extraction from limited user interactions. M-$LLM^3$REC comprises three integrated modules: the Motivation-Oriented Profile Extractor (MOPE), Motivation-Oriented Trait Encoder (MOTE), and Motivational Alignment Recommender (MAR). By emphasizing motivation-driven semantic modeling, M-$LLM^3$REC demonstrates robust, personalized, and generalizable recommendations, particularly boosting performance in cold-start situations in comparison with the state-of-the-art frameworks.

\end{abstract}




\maketitle

\section{Introduction}
In contemporary digital platforms, recommendation systems serve as critical intermediaries connecting users with information, products, and services~\cite{app10217748}, becoming essential tools for enhancing user experience and platform efficiency. These systems not only help users quickly identify content of interest but also significantly reduce decision-making costs under conditions of information overload~\cite{ISINKAYE2015261}. Therefore, constructing efficient and intelligent recommendation systems is of paramount importance for improving user engagement and platform operational capabilities~\cite{10.1145/3370082}.

Over the years, recommendation system technologies have continuously evolved, giving rise to several widely adopted mainstream approaches. Early methods, such as rule-based and content-based filtering, relied heavily on users' historical records to recommend items based on content similarity~\cite{Pazzani2007,cite-key}. Collaborative filtering later emerged as a powerful approach~\cite{10.1145/245108.245126,10.1145/371920.372071}, exploiting user-item interaction relationships to uncover behavioral similarities among users, thereby achieving more effective personalized recommendations. Matrix factorization methods presented novel solutions to address the latent challenges, such as the data sparsity issues~\cite{5197422}. More recently, with deep learning, neural collaborative filtering (NCF)~\cite{10.1145/3038912.3052569} and sequential modeling methods based on RNNs and Transformers~\cite{hidasi2016sessionbasedrecommendationsrecurrentneural,kang2018selfattentivesequentialrecommendation} have been introduced, aiming to capture more intricate user preferences and item relationships, thereby substantially improving recommendation accuracy and relevance.

Despite the notable progress in improving recommendation accuracy, these mainstream methods still face substantial challenges in sparse-data scenarios such as cold-start and long-tail users~\cite{10.1145/564376.564421,10.1145/1352793.1352837}. These challenges arise primarily because new users or niche items lack sufficient interaction records, making it challenging for models to learn reliable preference patterns~\cite{BOBADILLA2013109,10.1145/3357384.3357895}. Thus, the recommendation results based on extremely short behavioral sequences are often unstable and exhibit limited generalisability~\cite{10.1145/2043932.2043957}.

Recent studies have endeavored to mitigate cold-start issues by integrating cutting-edge technologies, which can be broadly categorized into two representative approaches~\cite{BOBADILLA2013109,Zhang_2020}. The first category, exemplified by methods like DiffuASR~\cite{BOBADILLA2013109} and Diff4Rec~\cite{10.1145/3581783.3612709}, exploits the generative capability of diffusion models to fill data gaps~\cite{10.1145/3539618.3591663} for cold-start users by generating pseudo-interaction sequences. Although these approaches partially extend user behavior sequences, it presents two fundamental limitations. First, the original cold-start sequences are inherently very short and contain limited information. As such, the generated pseudo-sequences often repeat existing signals rather than providing genuinely valuable semantic insights~\cite{liu2023llmrecbenchmarkinglargelanguage}. Second, the stochastic nature of diffusion processes can introduce noise into the generated samples, potentially resulting in misleading signals that compromise recommendation stability and precision~\cite{nichol2021improveddenoisingdiffusionprobabilistic,ho2020denoisingdiffusionprobabilisticmodels}.

The second category leverages product semantics to enrich user representations~\cite{cui2022m6recgenerativepretrainedlanguage,geng2023recommendationlanguageprocessingrlp}. These models, such as LLM-ESR~\cite{liu2024llmesrlargelanguagemodels} and SeRALM, utilize large language models (LLMs) to model product descriptions~\cite{10.1145/3626772.3657782}, recommending items based on semantic similarity between historical and candidate items. However, this strategy implicitly assumes consistent user interest in similar products, neglecting potential shifts in user motivation, which are common in real world~\cite{10.1145/3292500.3330989,ma2019hierarchicalgatingnetworkssequential}. In practice, user purchasing behaviors are frequently driven by dynamic intentions~\cite{mcauley2015inferringnetworkssubstitutablecomplementary}, and interests and needs are rarely static. Thus, recommendations that solely rely on historical semantic information risk being shortsighted and can't provide effective personalization~\cite{Zhang_2020}.

In light of these observations, we argue that effectively addressing cold-start problems hinges not primarily on behavioral sequence length, but rather on the ability to accurately extract deep motivational insights from limited interactions. Even when user activity on a platform is scarce, product texts, reviews, and purchase contexts encountered by users may still contain rich cues~\cite{10.1145/2507157.2507163,ni-etal-2019-justifying} reflecting their underlying purchasing motivations. Compared to superficial behavioral patterns, user motivations exhibit greater generalisability and transferability across contexts. Accurately modeling these motivations enables the constructions of robust user preference representations, even without extensive historical records, thereby facilitating more accurate predictions of future interests.

This perspective allows us to bypass the reliance on dense interaction histories by anchoring recommendations in transferable motivational semantics. It provides a practical shift enhancing both generalisability and robustness in cold-start scenarios. Motivated by this insight, we propose the first recommendation framework grounded in motivation-aware modeling, termed \textbf{M-$LLM^3$REC}. This framework comprises three integrated modules:
\begin{itemize}
    \item \textbf{Motivation-Oriented Profile Extractor (MOPE)}: We extract the semantically rich motivational representations from limited interaction traces of cold-start users. We further leverage the associated product descriptions and contextual information to infer the users' underlying intentions.
    \item \textbf{Motivation-Oriented Trait Encoder (MOTE)}: We abstract the items into interpretable motivational traits aligned with common motivation scopes. These traits representations serve as transferable semantic anchors, enabling cross-domain recommendation even without prior user-item interactions.
    \item \textbf{Motivational Alignment Recommender (MAR)}: We align the user motivational profiles with the candidate item traits, grounding recommendations in motivation-to-trait matching rather than interaction frequency. The alignment yields the generation of highly personalized recommendations, which are not only semantically consistent but also behaviorally plausible, especially for cold-start and sparse-data scenarios where users lack historical records. 
\end{itemize}

In summary, the key contributions of this work are as follows:
\begin{itemize}
    \item \textbf{Motivation-Aware Cold-Start Recommendation Paradigm.} We introduce the first recommendation framework that grounds cold-start user modeling in \textit{motivation-aware semantics}, departing from traditional behavior-centric paradigms. Unlike existing approaches such as DiffuASR~\cite{BOBADILLA2013109} and LLM-ESR~\cite{liu2024llmesrlargelanguagemodels}, which focus either on augmenting behavioral length or static semantic similarity, our work redefines cold-start modeling as a problem of uncovering \textit{generalizable motivational intent} from minimal user interactions.

    \item \textbf{Modular Architecture for End-to-End Motivation Modeling.} Our proposed framework, \textit{M-$LLM^3$REC}, is structured with three interdependent yet functionally distinct modules, which together form a closed-loop pipeline. This design captures the full life cycle of motivation-aware recommendation: from extracting implicit user intent (MOPE), to encoding item traits within a motivationally interpretable space (MOTE), to aligning user motivations with item semantics for final ranking (MAR). Such a modular decomposition enhances the model's transparency, extensibility, and control over recommendation pathways.

    \item \textbf{Motivation-to-Trait Alignment for Robust and Explainable Ranking.} We devise a novel \textit{motivation-to-trait alignment mechanism} that anchors the recommendation logic in semantic compatibility, rather than surface-level similarity heuristics. By grounding user-item matching in high-level motivational scopes, our framework enhances robustness under sparse data and significantly improves interpretability—offering not only \textit{what} to recommend but also \textit{why} it aligns with user intent.

    \item \textbf{Empirical Validation Across Cold-Start and Long-Tail Scenarios.} Through comprehensive evaluations on multiple benchmark datasets, we demonstrate that \textit{M-$LLM^3$REC} consistently outperforms state-of-the-art baselines, particularly in \textit{cold-start}, \textit{long-tail}, and \textit{zero-interaction} settings. These results substantiate the practical value of deep motivation modeling as a reliable and scalable solution for real-world recommendation systems facing data sparsity.
\end{itemize}

\section{Related Works}

In recent years, research on recommender systems has evolved along three major directions: traditional collaborative modeling methods, semantic modeling approaches based on large language models (LLMs), and interaction augmentation methods based on generative models.

Traditional methods primarily rely on historical user-item interactions for modeling. Classical collaborative filtering approaches such as NCF learn static user preferences~\cite{10.1145/3038912.3052569,kang2018selfattentivesequentialrecommendation}, while sequence-based models like SASRec capture user interest evolution by modeling behavioral sequences. However, these methods heavily depend on sufficient interaction records~\cite{10.1145/564376.564421,10.1145/3357384.3357895}, leading to significant performance degradation in cold-start and long-tail scenarios.

To address data sparsity, recent studies have explored incorporating LLMs to enhance semantic understanding~\cite{wu2024surveylargelanguagemodels} in recommendation systems. For instance, GenRec directly generates recommendation results based on item textual descriptions~\cite{cao2024genrecgenerativesequentialrecommendation}; LLMRec and LLM-ESR integrate graph structures or multimodal information to improve sparse relation modeling; P5 reformulates recommendation tasks as natural language generation, enhancing generalisability and adaptability to multi-task scenarios~\cite{liu2024llmesrlargelanguagemodels}. Despite their impressive performance, most of these approaches still focus on surface-level semantic similarity and lack the capacity to capture deeper user motivational intent behind behaviors~\cite{gao2023chatrecinteractiveexplainablellmsaugmented}.

Another line of research aims to augment training data by generating pseudo-interaction sequences~\cite{deldjoo2024reviewmodernrecommendersystems}. DiffuASR leverages a pretrained sequential recommender to provide gradient signals that guide a diffusion model in generating virtual item sequences~\cite{liu2023diffusionaugmentationsequentialrecommendation}. Diff4Rec introduces curriculum scheduling to gradually integrate generated sequences into training based on their complexity, mitigating noise interference~\cite{10.1145/3581783.3612709}. While such methods alleviate behavioral sparsity to some extent, the generated sequences often lack structurally grounded motivational semantics, resulting in limited generalization and controllability.

In summary, although existing methods have achieved progress in behavioral enhancement or semantic matching, there remains a significant gap in constructing a complete mechanism that models user motivational semantics from limited behaviors and drives recommendation accordingly. Our proposed framework, M-$LLM^3$REC, addresses this gap by systematically constructing motivation-based user profiling, trait abstraction, and matching strategies—achieving improved robustness, generalisability, and interpretability in challenging scenarios such as cold-start.

\section{Methodology}

In contrast to conventional behavior-centric recommender systems, which rely heavily on historical interactions, \textbf{M-$LLM^3$REC} is grounded in the hypothesis that user intent, rather than interaction density, is the true anchor of generalizable recommendation. To operationalize this perspective, we design a modular recommendation framework that models cold-start preferences via motivation-oriented semantic abstraction and alignment.

\begin{figure*}[t]  
    \centering
    \includegraphics[width=\textwidth]{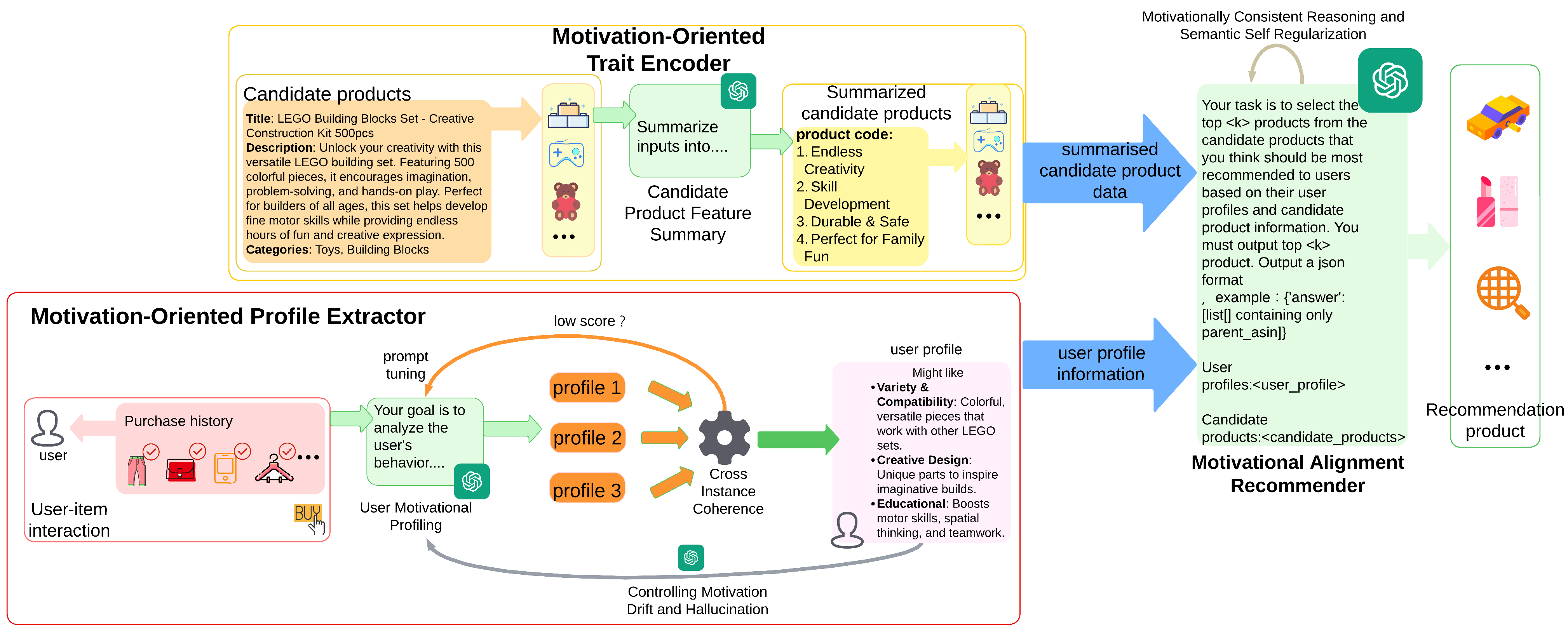}
    \caption{The figure illustrates the overall framework of the recommendation model. The top block represents the Motivational Alignment Recommender (MAR) module, followed by the Motivation-Oriented Profile Extractor (MOPE) module. Once the model generates the user motivation and extracts features from the 30 candidate products, both the user motivation and the processed candidate products are fed into the MAR module. Finally, the Motivational Alignment Recommender (MAR) outputs the top-k recommended items.
}
    \label{fig:wide}
\end{figure*}

The framework shown in Figure~\ref{fig:wide} comprises three interlinked modules, each addressing a core semantic transformation task: 1) Motivation-Oriented Profile Extractor constructs user intent representations with minimal behavioral records, learning implicit motivational intent from sparse interactions; 2) Motivation-Oriented Trait Encoder projects item semantics into an interpretable trait space aligned with motivational intent; and 3) Motivational Alignment Recommender grounds the final ranking in compatibility between inferred motivations and item-level traits, replacing traditional similarity-based heuristics with a conceptually aligned, intent-aware matching mechanism.

Together, these modules form an end-to-end, motivation-aware recommendation pipeline that translates sparse interactions into rich, intent-consistent representations, enabling robustness and interpretability in cold-start scenarios.

\begin{figure}[t]  
    \centering
    \includegraphics[width=0.23\textwidth]{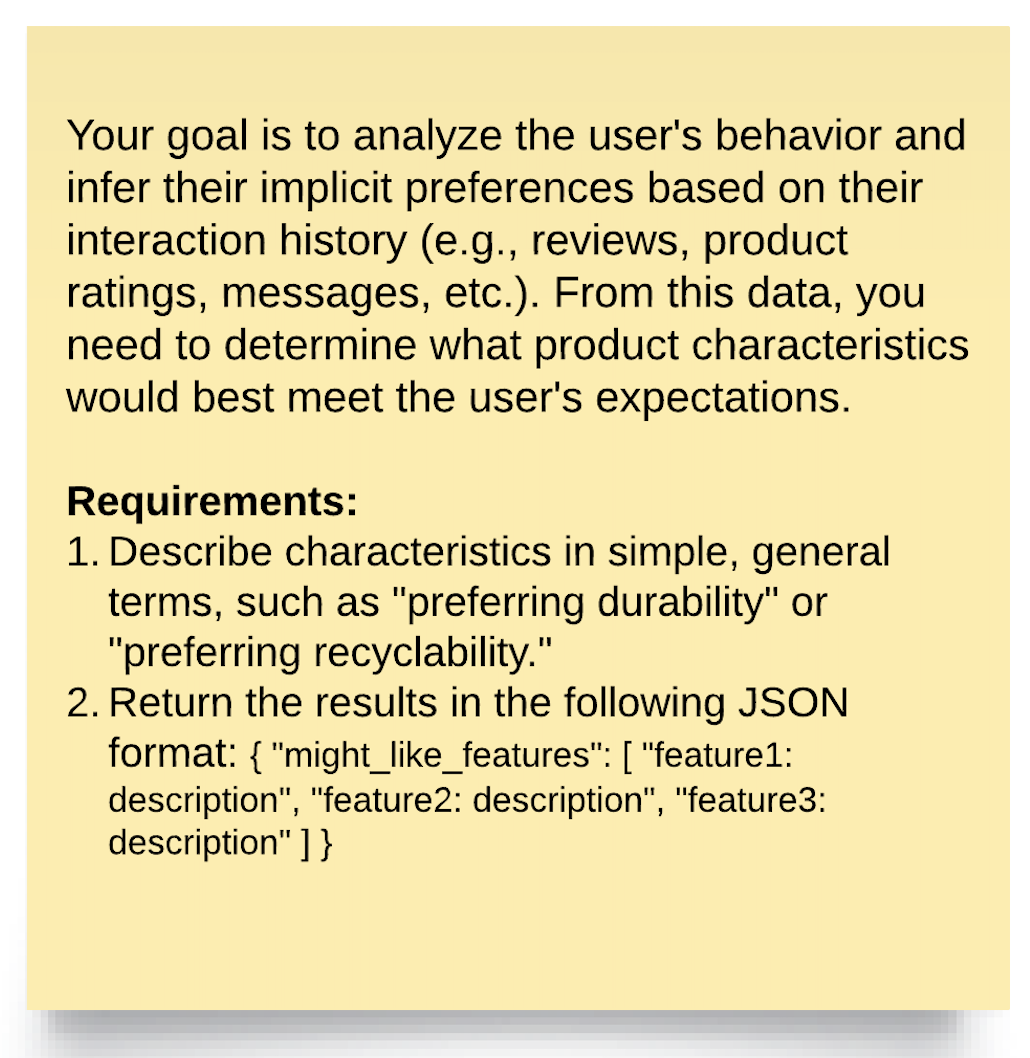}
    \includegraphics[width=0.23\textwidth]{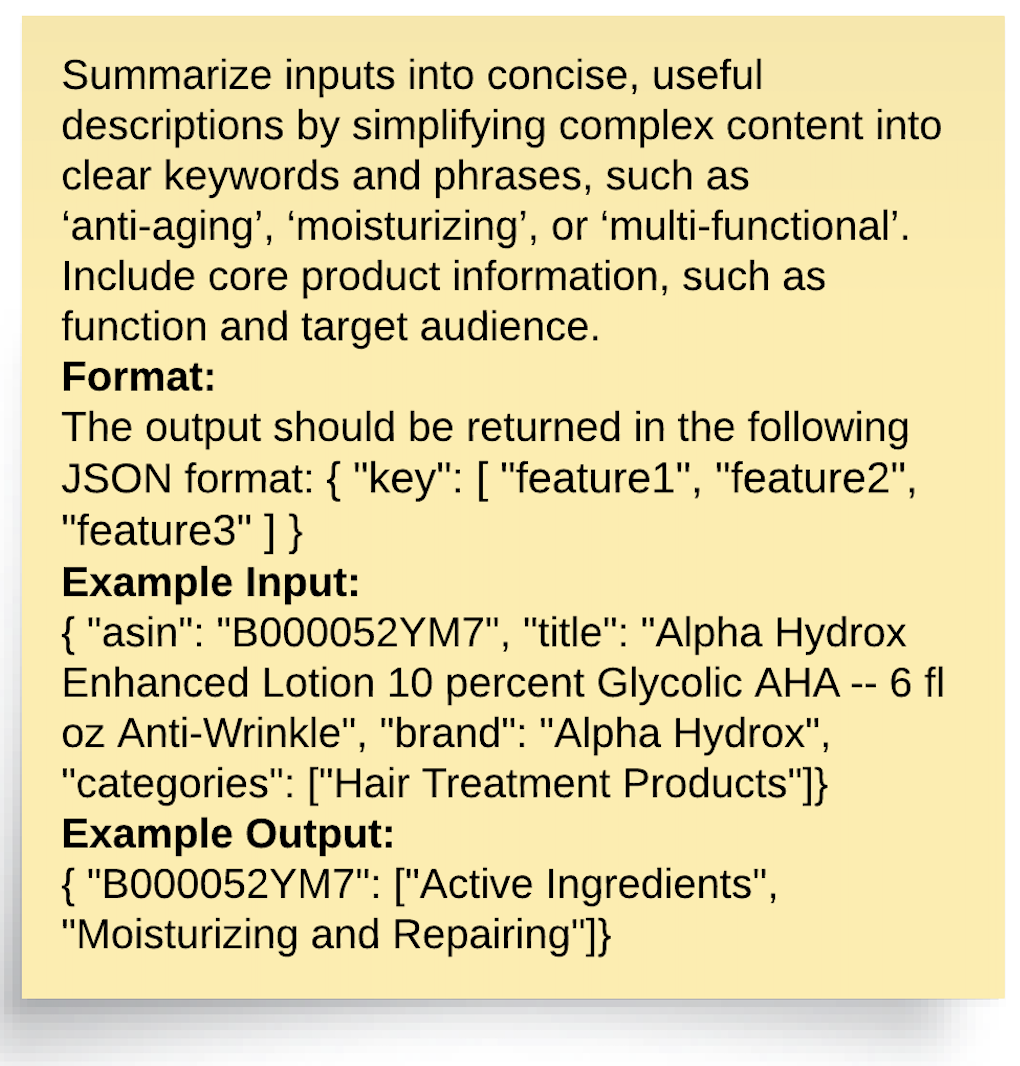}
    \caption{The prompt for MOPE module (left) and MOTE module (right).}
    \label{fig:wide1}
\end{figure}

\subsection{Motivation-Oriented Profile Extractor (MOPE)}

\subsubsection{Problem Formulation and Motivational Induction}

In cold-start settings, users present limited interaction histories, making it challenging to infer reliable preferences using behavior-based models. However, we observe that even sparse behavioral signals—when coupled with product semantics—can contain implicit motivational cues.

Let the interaction history of user \( u \) be represented as:
\begin{equation}
\mathcal{H}_u = \{(i_j, r_j)\}_{j=1}^k
\end{equation}
where \( i_j \) is an interacted item and \( r_j \) denotes the interaction signal (e.g., click, rating, or purchase). Each item is associated with a product description \( t_{i_j} \). We define the semantic context of user \( u \) as:
\begin{equation}
C_u = \texttt{Concat}(\{t_{i_j}, r_j\}_{j=1}^k)
\end{equation}
which concatenates descriptions and feedback into a single textual input sequence.

The goal of MOPE is to infer a structured motivational profile:
\begin{equation}
\bm{m}_u = \mathcal{F}_{\text{LLM}}(C_u; P_{\text{motivation}})
\end{equation}
where \( \mathcal{F}_{\text{LLM}} \) denotes the inference performed by a pretrained language model under prompt \( P_{\text{motivation}} \). The output \( \bm{m}_u \) is a structured set:
\begin{equation}
\bm{m}_u = \{ f_i : d_i \}_{i=1}^d
\end{equation}
where \( f_i \in \mathcal{M} \), and \( \mathcal{M} \) is a predefined motivational schema including dimensions such as \textit{functionality}, \textit{aesthetic}, and \textit{sustainability}.

This formulation frames the generation task as a semantic abstraction process that generalizes intent from behaviorally sparse data.

\subsubsection{Schema-Constrained Prompting and Generative Reasoning}

MOPE leverages In-Context Learning by conditioning the LLM on motivation-aware prompt templates containing annotated examples of behavior-to-intent mappings. The model generates motivational profiles aligned with schema categories, emphasizing semantic abstraction over item surface details. This structured representation supports interpretability and enables downstream alignment with item-level traits.

\subsubsection{Controlling Motivational Drift and Hallucination}

Due to the open-ended nature of LLMs, semantic drift and hallucination can occur. We mitigate this via:

\textbf{(1) Motivational Schema Constraint:} \\
By constraining outputs within a schema space \( \mathcal{M} \), we regulate semantic variance and encourage interpretable generation across functional, emotional, and ethical dimensions.

\textbf{(2) Reflective Prompting:} \\
An introspective generation mechanism asks the model to reassess prior motivations:

\begin{tcolorbox}[
    colback=gray!10,    
    colframe=gray!80,   
    boxrule=0.5pt,
    arc=2mm,
    fontupper=\ttfamily\small,
    breakable
]
``You previously inferred: X. Do you still agree based on the full history?''
\end{tcolorbox}

This reinforces coherence and helps eliminate implausible outputs.

\subsubsection{Cross-Instance Semantic Coherence}

To ensure semantic consistency across motivational explanations derived from multiple interactions, we design a structure-aligned coherence strategy. Given a user with \( k \) historical interactions, the LLM produces a set of motivational profiles \( \{\hat{m}_u^{(j)}\}_{j=1}^k \). We define a pairwise coherence score:
\begin{equation}
\text{Consistency}(\{ \hat{m}_u^{(j)} \}) = \frac{1}{k^2} \sum_{p=1}^k \sum_{q=1}^k \text{Sim}(\hat{m}_u^{(p)}, \hat{m}_u^{(q)})
\end{equation}
where \( \text{Sim}(\cdot, \cdot) \) can denote schema co-occurrence, keyword overlap, or semantic similarity of motivation descriptors.

We leverage the final score as a \textit{heuristic signal for prompt tuning}. In practice, when motivational outputs diverge significantly across samples, such as one emphasizing ``eco-consciousness'' and another focusing on ``premium aesthetic'', we will adjust the motivational prompt templates to encourage more focused and semantically aligned generations. Adjustments include modifying examples, clarifying category definitions, or emphasizing dominant behavioral cues.

This lightweight calibration mechanism plays a crucial role under sparse and noisy conditions, improving the reliability and interpretability of motivational induction in real-world cold-start settings.

\subsubsection{Personalization via Metadata Conditioning}

Different users may share the same interaction but be driven by distinct motivations. For instance, one user may purchase a product for its functionality, another for its aesthetic appeal. To capture this motivational heterogeneity, we introduce user metadata \( z_u \in \mathbb{R}^m \) (e.g., age, gender, region) into the generation process:
\begin{equation}
\bm{m}_u = \mathcal{F}_{\text{LLM}}(z_u, C_u; P_{\text{motivation}})
\end{equation}
This enables the LLM to produce motivational profiles that are demographically aligned and contextually relevant.

\subsection{Motivation-Oriented Trait Encoder (MOTE)}

\subsubsection{Motivation-Aligned Trait Abstraction}
In the context of this framework, \textit{alignment} refers to the conceptual consistency between the extracted product traits and the user's inferred motivational intent. Rather than relying on numerical vector similarity or co-interaction patterns, we define alignment as the semantic coherence between two sets of textual abstractions—user motivations and product traits. This alignment is achieved through carefully constructed prompts that guide the LLM to distill traits reflecting not surface attributes but the underlying affordances of the product. The goal is to ensure that the traits encapsulate generalized, motivation-relevant concepts (e.g., ``ease of use'', ``eco-friendliness''), which can be directly interpreted and matched against user motivations. This prompt-based conceptual alignment reduces semantic noise and ambiguity, enabling more robust and intention-consistent recommendations, especially in sparse-data or cold-start scenarios.

To align item representations with user motivations, we abstract each product into a trait-based semantic form that emphasizes interpretability and generalisability. Raw product descriptions are often verbose and domain-specific, which hinders direct comparison. The MOTE module addresses this by distilling each item's descriptive content into a structured set of high-level \textit{traits} that reflect its functional purposes, user applicability, and motivational relevance.

Given an item \( i \) with associated textual description \( d_i \), we define a trait extraction function:
\begin{equation}
\mathcal{T}_i = \mathcal{G}_{\text{LLM}}(d_i; P_{\text{trait}}),
\end{equation}
where \( \mathcal{T}_i = \{ t_1, t_2, \ldots, t_m \} \) denotes the set of distilled traits, and \( \mathcal{G}_{\text{LLM}} \) is a generative function guided by a structured prompt \( P_{\text{trait}} \). We assume each trait \( t_k \in \mathcal{V} \), where \( \mathcal{V} \) is a vocabulary of interpretable semantic descriptors such as \textit{``durability''}, \textit{``suitable for sensitive skin''}, or \textit{``easy to clean''}. Traits capture functional purposes, applicability conditions, and abstract selling points—rather than superficial keywords.

\subsubsection{Prompted Semantic Distillation}

To avoid the generation of trivial or surface-level features (e.g., brand names, pricing), the LLM is conditioned on carefully designed prompts that emphasize abstraction and generality. Specifically, prompts guide the model to extract traits that satisfy the following principles:

\begin{itemize}
    \item \textbf{generalisability:} Traits should describe transferable properties (e.g., ``supports joint health'') rather than transient specifications.
    \item \textbf{Functionality over Form:} Emphasize what the product does, not how it looks or is packaged.
    \item \textbf{Semantic Conciseness:} Each trait is distilled into a short phrase, enabling direct matching with motivational cues.
\end{itemize}

This setup encourages the LLM to act as a \textit{semantic expert} that summarizes high-level product affordances, akin to a well-informed salesperson highlighting what matters to prospective buyers.

\noindent \textit{Trait Representation and Downstream Alignment:} The traits \( \mathcal{T}_i \) serve as interpretable anchors for semantic alignment with user motivational profiles. Since traits are textual and categorical by nature, they enable alignment without requiring numerical historical co-interaction data. This supports generalization to new items and enhances transparency in recommendation logic.

\subsection{Motivational Alignment Recommender (MAR)}

\subsubsection{Intent-to-Trait Compatibility Modeling}

Having extracted user motivational profiles and item-level semantic traits, the core challenge becomes aligning these representations in a goal-directed, intent-aware manner. Inspired by alignment-based retrieval and zero-shot ranking frameworks, we formalize the motivational alignment as a non-parametric compatibility function:

\begin{equation}
\text{Score}(u, i) = \mathcal{F}_{\text{LLM}}(\bm{m}_u, \mathcal{T}_i),
\end{equation}

where \( \bm{m}_u \) denotes the motivational profile of user \( u \), and \( \mathcal{T}_i \) represents the distilled semantic trait set of item \( i \). The function \( \mathcal{F}_{\text{LLM}} \) is realized through a structured prompt passed to a pretrained language model, which infers the semantic compatibility between motivations and item traits.

This approach explicitly avoids reliance on surface-level similarity heuristics such as keyword overlap or co-interaction frequency. Instead, it focuses on aligning user intent with deep, abstracted product traits—allowing for more meaningful, generalizable recommendations even in sparse or cold-start contexts.

\subsubsection{Prompt-Based Ranking via Semantic Inference}

Instead of learning explicit parametric scorers, MAR leverages prompt-based generative alignment. For each candidate item \( i \in \mathcal{A}_u \), we instantiate an alignment query combining \( \bm{m}_u \) and \( \mathcal{T}_i \), prompting the LLM to determine their compatibility.

Let \( \mathcal{I} \) denote the global item pool available to the recommender system, and let \( \mathcal{A}_u \subset \mathcal{I} \) be the candidate set retrieved for user \( u \). Each item \( i \in \mathcal{A}_u \) is independently evaluated for motivational compatibility. The final Top-K recommendation is defined as:

\begin{equation}
\text{TopK}_u = \arg\max_{i \in \mathcal{A}_u} \text{Score}(u, i),
\end{equation}

where \( \text{Score}(u, i) \) is the alignment function inferred by the LLM over the pair \( (\bm{m}_u, \mathcal{T}_i) \).

\subsubsection{Motivationally Consistent Reasoning and Semantic Self Regularization}

While the model inherently infers alignment from structured input, the stochasticity of generative reasoning introduces risks of semantic drift and hallucination. To enhance robustness, we integrate a self-regularizing reasoning protocol grounded in motivation consistency.

After the initial Top-K selection, the model is prompted to generate brief rationale statements for each chosen item, explicitly linking them to the user’s motivational profile. These rationales are then re-evaluated via a follow-up prompt to ensure epistemic fidelity—i.e., whether the justification remains faithful to the inferred intent space.

This two-stage reflection mechanism introduces a semantic consistency prior that filters out implausible or misaligned recommendations. It improves the traceability, reliability, and interpretability of MAR, all while requiring no additional supervision or architectural change.

\subsection{Prompt-Oriented Model Conditioning via In-Context Learning}

Large language models (LLMs) are highly sensitive to prompt formulation. To ensure robust and interpretable generation across our framework, we adopt a structured prompting strategy rooted in In-Context Learning (ICL) (Figure \ref{fig:wide1}).

Specifically, each module—MOPE, MOTE, and MAR—is guided by purpose-specific prompts designed to convey the structure of the desired output (e.g., motivation schema, trait phrases, alignment decision). These prompts contain natural language exemplars that teach the LLM how to perform task-specific semantic transformations.

By embedding illustrative context and constraining output formats (e.g., JSON-style key-value pairs, enumerated features), we improve generation fidelity and enable semantic consistency across user profiles, item traits, and alignment decisions. This design transforms prompting from a heuristic into a modular and reproducible conditioning protocol.

\section{Experiment}

\subsection{Dataset}

To thoroughly evaluate our proposed framework, we use three benchmark datasets from the 2018 Amazon dataset~\citep{ni-etal-2019-justifying}, namely \textit{Beauty}, \textit{Toys}, and \textit{Sports}. Table~\ref{tab:data_table} summarizes the number of users, products, and interactions in each dataset. Each interaction includes \texttt{reviewerID}, \texttt{asin} (product ID), rating, and product-specific textual information.

To ensure sufficient data for inference, we filtered out users with fewer than two interactions. For each remaining user, we selected one positive sample---the highest-rated product in their interaction history (with rating $>$ 3)---as the ground-truth target.Then, we sample 29 negative candidate items and use the last item in the user sequence with a rating above 3 as the positive item, forming a test set with a total of 30 items.

\begin{table}[h]
    \centering
        \begin{tabular}{l r r r}
            \hline
            & \#User  & \#Item   & \#Interaction \\ \hline
            Beauty   & 22264 & 493003 & 196358      \\ 
            Toys     & 19333 & 374439 & 149654      \\ 
            Sports   & 35539 & 640478 & 283037      \\  \hline
        \end{tabular}
    \caption{User, Item, and Interaction Data}
    \label{tab:data_table}
\end{table}

Since our approach relies on Large Language Models (LLMs) without parameter tuning, it operates in an inference-only manner and does not require a training set. All methods, including baselines, are evaluated using the same user-level test data.

We report performance using two standard metrics: Hit Rate@K (HR@K) and Normalized Discounted Cumulative Gain@K (NDCG@K). Both metrics are computed \textit{per user} and then averaged over all users, following standard protocol in recommendation evaluation.

\subsection{Implementation}
Our framework incorporates the capabilities of GPT as the underlying LLMs, with different model versions tailored to specific components of the system, which are \textbf{gpt-3.5-turbo}, \textbf{gpt-3.5-turbo}, and \textbf{gpt-4o}. The max tokens of the product summarize module is set to 4095, and the p value of the module is set to 0.9, the temperature is set to 0.9, and the p value of the module is set to 0.9. The user profile module has its p value set to 1.0 and temperature set to 1.0. The recommended module has its p value and temperature set to 0.9. It is worth mentioning that changes in the hyper parameter do not have an impact on the performance of the model

\subsection{Evaluation Setting} 
We adopt two widely recognized metrics to evaluate the performance of our framework: Normalised Discounted Cumulative Gain (NDCG@K) and Hit Rate@K (HR@K). These metrics are included to assess both the ranking quality of recommendations and the likelihood of users interacting with the top-ranked items. In this work, we set $K=5$ and $K=10$ to measure the performance across different recommendation depths.

\subsection{Baseline}
To ensure a comparative analysis, we benchmark our framework with several state-of-the-art models, including \textit{SASRec}~\citep{kang2018selfattentivesequentialrecommendation} and  \textit{BERT4Rec}~\cite{sun2019bert4recsequentialrecommendationbidirectional} for sequence recommendation model. \textit{BPR}~\citep{rendle2012bprbayesianpersonalizedranking}, \textit{LightFM}~\citep{DBLP:conf/recsys/Kula15} for collaborative filtering. \textit{P5}~\citep{geng2023recommendationlanguageprocessingrlp},\textit{Tallrec}~\cite{Bao_2023}, \textit{A-LLMRec}~\cite{kim2024largelanguagemodelsmeet}, \textit{LLM-only} for LLM. \textit{LLM-only} refers to a minimal baseline that directly uses a large language model (LLM) to perform recommendation without incorporating any intermediate modules (e.g., user profiling or product summarization). Specifically, the input to the LLM consists of the raw user interaction history (i.e., product titles and descriptions) and a set of candidate items with their full descriptions. The LLM is prompted to generate top-$K$ recommendations solely based on these raw texts, without using any structured feature extraction or semantic summarization. This baseline serves to measure the performance of a direct LLM application without any task-specific adaptation or optimization. For the other models, we briefly list the key methodologies:
\begin{itemize}
    \item \textit{SASRec}~\citep{kang2018selfattentivesequentialrecommendation}: a self-attentive sequential algorithm that captures user preferences through sequential patterns.
    \item \textit{BERT4Rec}~\cite{sun2019bert4recsequentialrecommendationbidirectional}: The bidirectional model using the Cloze task, predicting the masked items in the sequence by jointly conditioning on their left and right context
    \item \textit{BPR}~\citep{rendle2012bprbayesianpersonalizedranking}: a Bayesian Personalized Ranking method for collaborative filtering.
    \item
    \textit{LightFM}~\citep{DBLP:conf/recsys/Kula15}:a library of recommender systems combining collaborative filtering and content filtering for processing data with user and item characteristics.
    \item 
    \textit{P5}~\citep{geng2023recommendationlanguageprocessingrlp}: a cutting-edge model using pre-trained LLMs for recommendation tasks.
    \item 
    \textit{Tallrec}~\cite{Bao_2023}: Aligns large language models efficiently to recommendation tasks via a two-stage instruction tuning process (Alpaca fine-tuning + Rec-tuning), combined with lightweight LoRA adaptation.
    \item
    \textit{A-LLMRec}~\cite{kim2024largelanguagemodelsmeet}:Through an alignment network, the user and item embeddings from a pre-trained collaborative filtering model (CF-RecSys) are mapped into the token space of a large language model (LLM), enabling the LLM to effectively integrate collaborative knowledge with textual information. A-LLMRec is the state-of-the-art (SOTA) method in 2024.

\end{itemize}

\subsection{Performance comparison}

\begin{table*}[t]
\centering
\resizebox{\textwidth}{!}{
\begin{tabular}{l|cccc|cccc|cccc}
\hline
\textbf{} & \multicolumn{4}{c|}{\textbf{Beauty}} & \multicolumn{4}{c|}{\textbf{Sports}} & \multicolumn{4}{c}{\textbf{Toys}} \\
\textbf{} & \textbf{NDCG@5} & \textbf{HR@5} & \textbf{NDCG@10} & \textbf{HR@10} & \textbf{NDCG@5} & \textbf{HR@5} & \textbf{NDCG@10} & \textbf{HR@10}  & \textbf{NDCG@5} & \textbf{HR@5} & \textbf{NDCG@10} & \textbf{HR@10}\\
\hline
BPR & 0.1084 & 0.1820 & 0.1654 & 0.3597	& 0.0863 & 0.1496 &	0.1418	& 0.3252 &	0.0902 & 0.1546 & 0.1434 & 0.3224\\
LightFM & 0.2475 & 0.3261 & 0.2819 & 0.4302 & 0.1205 & 0.2132 & 0.2587 & 0.3346 & 0.2089 & 0.2913 & 0.2914 & 0.3283 \\
\hline
SASREC & 0.2735 & 0.3572 & 0.3113 & 0.4604 & 0.2365 & 0.3285 & 0.2732 & 0.449 & 0.277 & 0.3574 & 0.3093 & 0.4611 \\
BERT4Rec & 0.1629 & 0.2733 & 0.2523 & 0.4949 & 0.1948 & 0.2975 & 0.2499 & 0.4729 & 0.192 & 0.2996 & 0.2376 & 0.4725\\
\hline
P5 & 0.1673 & 0.2448 & 0.1993 & 0.3441 & 0.1483  & 0.2241 & 0.1827 & 0.3313 & 0.0916 & 0.1411 & 0.1178 & 0.2227\\
LLM-only & 0.016 & 0.15 & 0.0795 & 0.25 & 0.082 & 0.17 & 0.135 & 0.32 & 0.093 & 0.2 & 0.113 & 0.37\\
Tallrec & 0.2097 & 0.32 & 0.255 & 0.5252 & 0.2122 & 0.35 & 0.2907 & 0.56 & 0.2659 & 0.3961 & 0.3271 & 0.6091\\
A-LLMRec & \textbf{0.4915} & 0.5365 & \textbf{0.4927} & 0.5307 & 0.1394 & 0.2180 & 0.200 & 0.3546 & 0.2496 & 0.3210 & 0.3027 & 0.4360\\
\hline
\textbf{M-$LLM^3$REC} & 0.339 & \textbf{0.540} & 0.358 & \textbf{0.709} & \textbf{0.2387} & \textbf{0.3876} & \textbf{0.3060} & \textbf{0.5734} &\textbf{0.2890} & \textbf{0.488} & \textbf{0.3518} & \textbf{0.684} \\
\hline
\end{tabular}
}
\caption{This table shows the experimental results of M-$LLM^3$REC compared with all baseline models. To ensure the accuracy of the experiments, we conducted three performance evaluations for each model and reported their average values.}
\label{tab:comparison}
\end{table*}

\subsubsection{Overall Model Performance Comparison}

As shown in Table~\ref{tab:comparison}, M-$LLM^3$REC demonstrates outstanding performance across all metrics (NDCG@5, HR@5, NDCG@10, HR@10) on the Beauty, Sports, and Toys datasets, with a clearly superior overall capability compared to all baseline models. Particularly for HR@10, a key metric reflecting recommendation coverage and hit rate, M-$LLM^3$REC achieves the highest scores across all three domains: Beauty (0.709), Sports (0.5734), and Toys (0.684), significantly outperforming other models. The leading performance on NDCG metrics also indicates that the model excels in ranking quality, effectively placing the most relevant items at the top of the recommendation list.

\subsubsection{Comparison with Sequential Modeling Models}

SASRec and BERT4Rec are both sequence-based recommendation models with advantages in modeling the order of user behaviors. On the Beauty dataset, SASRec achieves an HR@10 of 0.4604, and BERT4Rec reaches 0.4949, while M-$LLM^3$REC achieves 0.709, reflecting improvements of 54\% and 43\%, respectively. In Sports, SASRec and BERT4Rec reach 0.449 and 0.4729, while M-$LLM^3$REC boosts this to 0.5734, yielding over 20\% improvement.

These comparisons show that behavioral sequence modeling alone is insufficient to capture high-dimensional semantic preferences and fine-grained interests. In contrast, M-$LLM^3$REC, by introducing user motivation and semantic enhancement mechanisms, builds on behavioral data to model deeper user intent, substantially improving recommendation outcomes.

\subsubsection{Comparison with Traditional Models}

BPR and LightFM represent traditional recommendation approaches based on matrix factorization and hybrid features, respectively, yet both lag behind in performance across all datasets. For instance, on the Beauty dataset, BPR scores show 0.3597 and LightFM reaches 0.4302 in HR@10, compared to 0.709 for M-$LLM^3$REC. On Toys, LightFM achieves just 0.3284, and BPR lower at 0.3224, while M-$LLM^3$REC reaches 0.684, representing nearly an order of magnitude improvement.

This highlights the inadequacy of traditional models when dealing with high-dimensional, multi-category, and semantically complex recommendation scenarios, due to their lack of deep user understanding and nonlinear modeling capacity.

\subsubsection{Comparison with Language Model–Based Methods}

P5 and LLM-only are pure language model–based recommendation methods. While they offer some degree of semantic understanding, their lack of task-specific adaptation mechanisms leads to inferior performance. For example, on Toys, P5 and LLM-only achieve HR@10 scores of 0.2227 and 0.37, respectively—far below M-$LLM^3$REC’s 0.684.

Tallrec, which incorporates structured information modeling, shows competitive results on some datasets (e.g., HR@10 of 0.6091 on Toys), yet still falls short of M-$LLM^3$REC in terms of generalization. A-LLMRec, though performing well in Beauty (NDCG@5 = 0.4915, HR@10 = 0.5307), shows clear performance drops in Sports and Toys (HR@10 = 0.3506 and 0.4360, respectively). In contrast, M-$LLM^3$REC maintains leading performance across all domains, reflecting its superior robustness and cross-domain adaptability.

\subsubsection{Summary}

Overall, M-$LLM^3$REC outperforms traditional, sequential, and language model–based baselines, demonstrating stronger capabilities in user interest modeling, domain generalization, and recommendation ranking accuracy. Its integration of user motivation and semantic enrichment confirms the immense potential of combining structured personalization with language understanding in next-generation recommendation systems.

\begin{table}[htbp]
\centering
\resizebox{\linewidth}{!}{
\begin{tabular}{l|cc|cc|cc}
\hline
\textbf{} & \multicolumn{2}{c|}{\textbf{Beauty}} & \multicolumn{2}{c|}{\textbf{Sports}} & \multicolumn{2}{c}{\textbf{Toys}} \\
\textbf{} & \textbf{HR@5} & \textbf{NDCG@5} & \textbf{HR@5} & \textbf{NDCG@5} & \textbf{HR@5} & \textbf{NDCG@5} \\
\hline
BERT4Rec & 0.0704 & 0.0499 & 0.0674 & 0.0303 & 0.1311 & 0.0895\\
SASRec & 0.1199 & 0.1022 & 0.1842 & 0.0796 & 0.1257 & 0.0698 \\
M-$LLM^3$REC & 0.435 & 0.2502 & 0.45 & 0.2717 & 0.4975 & 0.3078\\
\hline
\end{tabular}
}
\caption{Performance comparison under the item cold-start setting.}
\label{tab:colditem}
\end{table}

\begin{table}[htbp]
\centering
\resizebox{\linewidth}{!}{
\begin{tabular}{l|cc|cc|cc}
\hline
\textbf{} & \multicolumn{2}{c|}{\textbf{Beauty}} & \multicolumn{2}{c|}{\textbf{Sports}} & \multicolumn{2}{c}{\textbf{Toys}} \\
\textbf{} & \textbf{HR@5} & \textbf{NDCG@5} & \textbf{HR@5} & \textbf{NDCG@5} & \textbf{HR@5} & \textbf{NDCG@5} \\
\hline
BPR & 0.1127 & 0.0565 & 0.0776 & 0.0355 & 0.1258 & 0.0770\\
LightFM   & 0.1219 & 0.0611 & 0.1149 & 0.0576 & 0.1064 & 0.0458 \\
BERT4Rec & 0.2633 & 0.1429 & 0.2766 & 0.1778 & 0.2696 & 0.172\\
Tallrec & 0.308 & 0.1845 & 0.342 & 0.2017 & 0.3725 & 0.2351\\
SASRec & 0.3203 & 0.2032 & 0.2708 & 0.1721 & 0.333 & 0.2212 \\
M-$LLM^3$REC & 0.464 & 0.2704 & 0.37 & 0.212 & 0.456 & 0.262\\
\hline
\end{tabular}
}
\caption{Performance comparison under the user cold-start setting.}
\label{tab:coldstart}
\end{table}

\section{Cold-Start Experiment}

To systematically evaluate the model’s recommendation capability under both product cold-start and user cold-start conditions, we designed corresponding experimental settings. In the \textbf{product cold-start} scenario, we selected the 10\% of products with the fewest user interactions as the cold-start test set, and excluded users who had interacted with these products from the user training set to form the test set. These cold-start products were treated as positive samples during prediction. The experimental results show that traditional sequential recommendation models perform significantly worse under product cold-start conditions compared to user cold-start scenarios. This is because sequential models heavily rely on mining the relationships between already trained products and users during training. Once a product has not been seen by the model during training, the model tends to recommend familiar products during inference, and thus fails to effectively handle cold-start products. In contrast, our proposed M-$LLM^3$REC model focuses on the rich feature information carried by both products and users, making it easier to identify semantic similarities between users and unseen products, and thus remains largely unaffected under product cold-start conditions.

At the same time, to systematically evaluate the model’s performance under \textbf{user cold-start} conditions, we adopted a standard cold-start evaluation method. Specifically, we sorted all users based on their interaction timestamps, took the latest 10\% of interaction records as the test set, and retained only users with fewer than three interactions to simulate a realistic scenario where new users have limited early behavior. During model evaluation, these cold-start users were completely excluded from both training and prompt construction to ensure fairness and generalisability. The experimental results show that M-$LLM^3$REC significantly outperforms traditional collaborative filtering methods (such as BPR, LightFM) and deep learning-based sequential models (such as SASRec, BERT4Rec, Tallrec) across datasets including Beauty, Sports, and Toys, achieving absolute advantages in both HR@5 and NDCG@5 metrics.

Traditional collaborative filtering methods heavily rely on dense user–product interaction matrices to infer latent preferences. Once in cold-start scenarios, there was a significant decline in their performance. Deep learning-based sequential models also struggle to address the cold-start problem effectively due to overfitting or information loss caused by sparse interaction data. In contrast, M-$LLM^3$REC leverages large language models to mine semantic information from text, directly matching user preferences with product descriptions. It utilizes a motivation-based inference mechanism to infer semantically rich user and product representations without requiring fine-tuning on cold-start users. Furthermore, its multi-source information fusion mechanism, which integrates explicit information (such as product metadata and user attributes) with implicit behavioral signals, enables the system to perform efficient recommendations even with minimal interaction data.

In conclusion, M-$LLM^3$REC, through an effective semantic-enhanced recommendation paradigm, fully exploits the semantic modeling and generalization capabilities of large language models. It demonstrates outstanding recommendation performance in both product and user cold-start scenarios, highlighting its practical value and broad applicability in sparse data environments.

\subsection{Ablation studies}

\begin{table}[htbp]
\centering
\resizebox{\linewidth}{!}{
\begin{tabular}{l|cc|cc|cc}
\hline
\textbf{} & \multicolumn{2}{c|}{\textbf{Beauty}} & \multicolumn{2}{c|}{\textbf{Sports}} & \multicolumn{2}{c}{\textbf{Toys}} \\
\textbf{} & \textbf{HR@5} & \textbf{NDCG@5} & \textbf{HR@5} & \textbf{NDCG@5} & \textbf{HR@5} & \textbf{NDCG@5} \\
\hline
Without MOPE & 0.175 & 0.076 & 0.21 & 0.096 & 0.250 & 0.111\\
Without MOTE & 0.478 & 0.285 & 0.32 & 0.190 & 0.45 & 0.231 \\
\hline
\end{tabular}
}
\caption{Ablation studies of model.}
\label{tab:ablation}
\end{table}

\begin{table}[htbp]
\centering
\resizebox{\linewidth}{!}{
\begin{tabular}{l|cc|cc|cc}
\hline
\textbf{} & \multicolumn{2}{c|}{\textbf{Beauty}} & \multicolumn{2}{c|}{\textbf{Sports}} & \multicolumn{2}{c}{\textbf{Toys}} \\
\textbf{} & \textbf{HR@5} & \textbf{NDCG@5} & \textbf{HR@5} & \textbf{NDCG@5} & \textbf{HR@5} & \textbf{NDCG@5} \\
\hline
gpt-3.5-turbo & 0.379 & 0.223 & 0.34 & 0.209 & 0.32 & 0.216 \\
gpt-4-turbo   & 0.564 & 0.340 & 0.45 & 0.255 & 0.49 & 0.290 \\
gpt-4o        & 0.480 & 0.275 & 0.44 & 0.258 & 0.48 & 0.273 \\
\hline
\end{tabular}
}
\caption{The performance with different GPT models as LLMs.}
\label{tab:gpt-performance}
\end{table}

\begin{table}[htbp]
\centering
\scalebox{1}{
\begin{tabular}{l|c|c|c}
\hline
\textbf{} & \textbf{Beauty} & \textbf{Sports} & \textbf{Toys} \\
\hline
gpt-3.5-turbo & 0.008 & 0.009 & 0.013 \\
gpt-4-turbo   & 0.100 & 0.087 & 0.090 \\
gpt-4o        & 0.058 & 0.034 & 0.052 \\
4o \& 3.5     & 0.018 & 0.009 & 0.013 \\
\hline
\end{tabular}
}
\caption{Cost of M-$LLM^3$REC per user-item interaction.}
\label{tab:cost}
\end{table}

\begin{table}[htbp]
\centering
\resizebox{\linewidth}{!}{
\begin{tabular}{l|cc|cc|cc}
\hline
\textbf{} & \multicolumn{2}{c|}{\textbf{Beauty}} & \multicolumn{2}{c|}{\textbf{Sports}} & \multicolumn{2}{c}{\textbf{Toys}} \\
\textbf{} & \textbf{HR@5} & \textbf{NDCG@5} & \textbf{HR@5} & \textbf{NDCG@5} & \textbf{HR@5} & \textbf{NDCG@5} \\
\hline
with gpt-3.5-turbo & 0.4 & 0.234 & 0.458 & 0.313 & 0.46 & 0.284\\
with gpt-4o & 0.53 & 0.268 & 0.462 & 0.311 & 0.48 & 0.292\\
\hline
\end{tabular}
}
\caption{MOPE model with different LLMs}
\label{tab:DifProfile}
\end{table}

To better understand the contribution of each module in M-$LLM^3$REC to the overall model performance, we conducted an ablation study. Specifically, we removed the MOPE Module and MOTE Module separately and evaluated the model's performance on the Beauty, Sports, and Toys datasets using HR@5 and NDCG@5. The results are shown in Table \ref{tab:ablation}.

The experimental results indicate that the MOPE Module is a core component in enhancing recommendation performance. When this module was removed, model performance dropped significantly. For example, in the Beauty dataset, HR@5 decreased from 0.54 to 0.175, and NDCG@5 dropped from 0.339 to 0.076, representing a reduction of over 67\% and 77\%, respectively. Similar trends were observed in Toys (HR@5 dropped from 0.58 to 0.250) and Sports (from 0.37 to 0.21), demonstrating that the user motivation extract mechanism plays a critical role in constructing personalized interest representations and capturing fine-grained preferences.

In comparison, the MOTE Module had a smaller but still noticeable impact. When removed, the HR@5 on the Beauty dataset dropped to 0.478 (about an 11.5\% decrease), and NDCG@5 to 0.285, indicating that semantic information from product summaries contributes positively to user understanding. On the Sports dataset, the model still maintained relatively high performance without this module (HR@5 = 0.32, NDCG@5 = 0.190), suggesting that MOPE plays an even more crucial role in this domain. A moderate performance drop was also observed on Toys, but the model retained strong results (HR@5 = 0.45, NDCG@5 = 0.231).

It is worth noting that the MOTE module was originally designed for input optimization. To align user motivations with the semantic information of items, it is necessary to preprocess the raw item data in a way that maximizes the performance of MOPE. As a user’s interaction history grows, the original product text may become excessively long, which can lead to inefficiencies or instability in LLM inference. While this module is not a decisive factor for recommendation accuracy, it holds significant engineering value in real-world deployment by improving inference speed and maintaining model stability.

In summary, the MOPE Module is the key factor driving performance gains across all scenarios. Its ability to accurately model user preferences forms the foundation of recommendation effectiveness. Meanwhile, the MOTE Module helps condense semantic content while reducing input complexity. Together, these two components synergistically enhance both the effectiveness and practicality of the model. This is especially important in information-rich domains like Beauty and Toys, where full semantic understanding is crucial for accurate recommendation.

\subsection{Model selection}

To further enhance the practicality and deployability of the recommendation system, we integrated different versions of large language models (GPT-3.5-turbo, GPT-4-turbo, and GPT-4o) into the M-$LLM^3$REC framework and conducted a systematic analysis of the trade-off between recommendation performance and inference cost.

\subsubsection{Performance improves with model upgrades (Table \ref{tab:gpt-performance})}
As shown in Table \ref{tab:gpt-performance}, recommendation performance continuously improves as the LLM is upgraded from GPT-3.5-turbo to GPT-4-turbo and GPT-4o. For example, in the Beauty dataset, HR@5 increases from 0.379 (GPT-3.5) to 0.564 (GPT-4-turbo), and NDCG@5 increases from 0.223 to 0.340, indicating that upgrading the LLM version significantly enhances user interest modeling. Similar trends appear in the Sports and Toys datasets. For instance, in Toys, GPT-4o achieves HR@5 = 0.48, very close to the performance of GPT-4-turbo (0.49).

\subsubsection{Cost analysis reveals cost-performance trade-offs (Table \ref{tab:cost})}
However, performance improvements come with increased costs. As shown in Table \ref{tab:cost}, GPT-3.5-turbo has the lowest inference cost in all scenarios (e.g., \$0.008 for Beauty), while GPT-4-turbo has the highest cost (\$0.100 for Beauty). GPT-4o falls between the two and offers better cost-performance balance in most cases. For example, in the Sports dataset, GPT-4o's cost is \$0.034, much lower than GPT-4-turbo's \$0.087, while achieving nearly equivalent or even better performance.

\subsubsection{Model selection strategy for the MOPE module (Table \ref{tab:DifProfile})}
We further analyzed the impact of different LLMs on system performance when applied to the MOPE module (Table \ref{tab:DifProfile}). The results show that upgrading the MOPE module from GPT-3.5-turbo to GPT-4o does not significantly improve performance, and in some cases (e.g., Sports). This suggests that in a multi-module recommendation architecture, some modules are less dependent on LLM capabilities and can use lower-cost models without affecting overall performance.

\subsubsection{Hybrid deployment strategy: best trade-off between performance and cost}
Taking both performance and cost into account, we propose a hybrid deployment strategy to achieve optimal cost-effectiveness: use GPT-4o for the MAR module to ensure high accuracy; use GPT-3.5-turbo for supporting modules such as MOPE module to reduce overall invocation cost. This strategy keeps the total cost per user-item interaction at \$0.013, which is approximately 87\% lower than the full GPT-4-turbo deployment, while maintaining near-optimal performance. 

\subsection{Discussion}
\subsubsection{Prompt Sensitivity Analysis}
In designing the three modules, we adopted distinct prompt structures with the primary aim of enabling each module to clearly comprehend its designated role while mitigating the influence of hallucinations in the LLM (Figure~\ref{fig:wide1}). To further reinforce each module's understanding of its assigned task, we incorporated a structured prompting format~\cite{rath2025structuredpromptingfeedbackguidedreasoning}, combined with few-shot examples~\cite{brown2020languagemodelsfewshotlearners}. This design consideration focuses on maintaining clarity of function across modules and does not compromise the overall robustness of the framework.

\subsubsection{Computational Overhead and Trade-offs}
Although LLMs inherently introduce substantial computational overhead due to their complex reasoning processes, their powerful reasoning capabilities in the LLM-driven era have led to significant performance gains across diverse domains. Moreover, in our proposed framework, it does not require fine-tuning or additional training requirements. We have eliminated the resource costs typically associated with model training. Therefore, the computational overhead can be regarded as a reasonable trade-off for the performance gains.

\section{Conclusion}
In this work, we introduced M-$LLM^3$REC, a novel semantic-aware recommendation framework that leverages large language models to model user-item interactions through a motivation-centric paradigm. Unlike traditional behavior-based or semantic similarity approaches, our method centers on uncovering users' underlying motivational intents, enabling more robust and generalizable recommendations—especially in cold-start and sparse-data scenarios. By integrating three core modules—MOPE (for motivation profiling), MOTE (for semantic trait abstraction), and MAR (for alignment-based recommendation)—M-$LLM^3$REC forms a closed-loop system that models, aligns, and matches user intent with product semantics in an interpretable manner. Extensive experiments across multiple datasets confirm that our model outperforms state-of-the-art baselines in both recommendation accuracy and generalization ability. Furthermore, ablation studies and cost-performance analyses demonstrate the practicality and scalability of our framework. Overall, M-$LLM^3$REC provides a significant step forward in utilizing LLMs for personalized, motivation-aware recommendation in real-world, data-sparse environments.


\bibliographystyle{ACM-Reference-Format}
\balance
\bibliography{main}


\end{document}